\documentclass[twocolumn,showpacs,amsmath,amssymb,prb,superscriptaddress,floatfix,raggedfooter]{revtex4}

\usepackage{graphicx}% Include figure files
\usepackage{dcolumn}% Align table columns on decimal point
\usepackage{bm}% bold math

\usepackage{amssymb}
\usepackage{amsmath}
\usepackage{amsfonts}
\usepackage{amssymb}
\usepackage{graphicx}
\usepackage{hyperref}
\usepackage{color}
\usepackage{epsfig}

\usepackage{xcolor}
\definecolor{red}{rgb}{0.7,0,0}%darkred MIT
\definecolor{green}{rgb}{0.,0.35,0.}%darkgreen
\definecolor{blue}{rgb}{0.2,0.2,0.7} %beamer@blendedblue
\definecolor{black}{rgb}{0.15,0.15,.15}%not too black

\begin{document}

%\title{Exploring magnetism with constrained dipolar bosons}

\title{Homogeneous and inhomogeneous magnetic phases of constrained dipolar bosons}

\date{\today}
%find a way to fix this affiliation mess!!
\author{M. Dalmonte}
\email[electronic address: ]{marcello.dalmonte@bo.infn.it}
\affiliation{Dipartimento di Fisica dell'Universit\`a di Bologna, via Irnerio 46, 40126 Bologna, Italy}
\affiliation{INFN, Sezione di Bologna, via Irnerio 46, 40126 Bologna, Italy}
\affiliation{IQOQI and Institute
for Theoretical Physics, University of Innsbruck, A-6020 Innsbruck, Austria}
\author{M. Di Dio}
\email[electronic address: ]{mario.didio@gmail.com}
\affiliation{Dipartimento di Fisica dell'Universit\`a di Bologna, via Irnerio 46, 40126 Bologna, Italy}
\author{L. Barbiero}
\email[electronic address: ]{luca.barbiero@polito.it}
\affiliation{Dipartimento di Fisica del Politecnico, corso Duca degli Abruzzi, 24, 10129 Torino, Italy}
\author{F. Ortolani}
\email[electronic address: ]{ortolani@bo.infn.it}
\affiliation{Dipartimento di Fisica dell'Universit\`a di Bologna, via Irnerio 46, 40126 Bologna, Italy}
\affiliation{INFN, Sezione di Bologna, via Irnerio 46, 40126 Bologna, Italy}

\begin{abstract}
We study the emergence of several magnetic phases in dipolar bosonic gases subject to three-body loss mechanism employing numerical 
simulations based on the density matrix renormalization group(DMRG) algorithm. After mapping the original Hamiltonian in spin language, we find a strong parallelism between the bosonic theory and the spin-1 Heisenberg model with single ion anisotropy and long-range interactions. A rich phase diagram, including ferromagnetic, antiferromagnetic and non-local ordered phases, emerges in the one-dimensional case, and is preserved even in presence of a trapping potential.

\end{abstract}

%\pacs{34.20.-b, 71.10.Pm, 03.75.Lm(ok), 05.30.Jp}
%03.75 Hh ()
%05.10 Cc (dmrg section)**
%05.30 Jp (boson system)
%67.85 Hj (bosons in OL)  (principal pacs!!)
%75.10 Pq (magnetismo)**  
\maketitle
 
 \section{Introduction}\label{introduction}

Recent experimental advances in controlling ultracold gases of magnetic atoms\cite{pfau} and polar molecules\cite{Ni} have paved the way to the investigation of several quantum many-body phenomena\cite{pmreviews}. These setups naturally provide anisotropic, long-range dipolar interactions, which can be tuned and manipulated with high accuracy in order to access the physics of spin systems\cite{micheli} and Hubbard-like models\cite{pmreviews} loaded into optical lattices\cite{bloch_rv}. Considerable theoretical efforts have focused on one dimensional geometry, where non-local interactions play a fundamental role in stabilizing interesting phenomena such as supersolidity\cite{batrouni}, checkboard insulator\cite{burnell,dalmonte} and insulating phases characterized by non-local order parameters\cite{dallatorre,amico}.

Furthermore, dissipative processes have emerged as an additional, relevant source of interaction. Two-body losses have been successfully employed to engineer hard core interactions in molecular gases\cite{syassen}, thus leading to the stabilization of a Tonks-Giradeau gas, and three-body losses have been proposed as a dynamical source of three-body interaction\cite{daley} which stabilizes dimer-superfluidity\cite{daley, diehl}, color-superfluidity\cite{kantian} and Pfaffian-like states\cite{roncaglia,paredes} with ultracold atoms.
 
The aim of this paper is to investigate the interplay between local (two- and three-body) and non-local interactions in low dimensional systems of ultracold dipolar bosons. We focus on a one dimensional geometry, and find that the phase diagram of such systems strongly resembles that of the  spin-1 Heisenberg 
model with Ising-like and single ion anisotropy~\citep{affleck,tasaki, kennedy,sanctuary}, or $\lambda-D$ model, extensively studied in the past in the contest of 
one-dimensional spin chains.
We present numerical results on the phase 
diagram, which exhibits ferromagnetic, antiferromagnetic and hidden order phases, and finally discuss the stability of these phases in presence of a 
trapping potential and density fluctuations, as naturally present in cold atomic and molecular setups. Several of the magnetic phases discussed do not require strong dipolar interactions, 
and can thus be observed even with magnetic atoms, where dipolar interaction is usually much smaller than any other relevant energy scale\cite{pmreviews}. 

The paper is organized as follows: in Sec. \ref{model}, we describe the parallelism between constrained bosonic gases and spin systems and introduce the Hamiltonian, which is then investigated by DMRG simulations and strong coupling arguments in Sec. \ref{hom_section}. In Sec. \ref{trap} we extend the numerical simulations to the inhomogeneous case; finally, we draw our conclusions in Sec. \ref{conclusions}.

\section{Bosonic Hamiltonian and $\lambda$-D model}\label{model}

 Dipolar bosons confined in a one dimensional geometry and subject  to a deep optical lattice are generally described by the following Hamiltonian \cite{pmreviews}:
\begin{eqnarray}\label{boseham}
\mathcal{H}&=&-t\sum_{\langle i,j\rangle}b^{\dag}_{i}b_{j}+\frac{U}{2}\sum_{i}n_{i}\left(n_{i}-1\right)+\mu\sum_in_i+\nonumber\\
&+&k\sum_i(i-L/2)^2n_i +\Lambda\sum_{i<j}\frac{n_in_j}{(j-i)^3}.
\end{eqnarray}
Here, $b^{\dagger}_i, b_i, n_i$ are bosonic creation, annihilation and number operator at site $i$, the first line describes the standard Bose-Hubbard model, where $t$ is the hopping term between nearest neighbor sites and $U$ the onsite interaction, and the last line includes trapping and  long-distance dipolar potentials. The hopping coefficient $t$ varies with the depth of the underlying optical lattice, whereas $\Lambda$ can be tuned by varying the applied EC electric field; finally, the onsite interaction $U$ depends on the short-distance details of the interparticle interaction\cite{goral} and, for magnetic atoms, can be controlled by using Feshbach resonances\cite{pmreviews}.
The phase diagram of Eq. (\ref{boseham}) with $U>0$ has been investigated in several regimes: at unitary filling, a new insulating phase characterized by hidden order has been predicted between a Mott insulator and a charge density wave\cite{dallatorre, amico}, whereas for densities $\bar{n}<1$ and strong repulsive interaction $U\gg t$ a devil's staircase of insulating phases appears as a function of the chemical potential\cite{burnell}. 
However, the attractive regime $U<0$ has so far been neglected. This is partially due to the fact that losses given by strong three-body recombination are enhanced in this regime, thus making a time-dependent description of the system more suitable in order to take into account dissipative effects\cite{brennen}. 
The situation can be though strongly simplified when the decay rate $\gamma_3$\cite{daley} associated with three-body loss processes is much larger than the typical tunneling rate, i.e. $\gamma_3\gg t$: in this regime, a mechanism analogous to the quantum Zeno effect gives rise to an effective strong three-body repulsion, which can be implemented in the Hamiltonian with the additional condition $(b^{\dagger}_i)^3=0$\cite{daley}. 

The opportunity to engineer strong three-body repulsion has then two striking effects: {\it i)} the system is in general stable regardless of the sign of the couplings $U,\Lambda$ and {\it ii)} the onsite Hilbert space is reduced to $|0\rangle, |1\rangle,|2\rangle$, thus resembling a spin-1 system. This correspondence is further clarified after introducing spin-1 operators $S_i^{+}, S_i^-, S_i^z$ and performing the following mapping:
\begin{equation}
n_i=1-S_i^z, \qquad b^{\dagger}_i=\alpha S_i^-+\beta(S^z_iS^-_i+S^-_iS^z_i)
\end{equation}
where $\alpha=(2+\sqrt{2})/4, \; \beta=-(2-\sqrt{2})/4$ are fixed by commutation relations, as described in appendix \ref{app_map}. From now on, we will consider a fixed density $\bar{n}=1$, then obtaining (fixing $t=1$):
\begin{eqnarray}\label{hamspin}
\mathcal{H}&=& -\sum_{<i,j>}S^+_iS^-_{j}(J+J_1S^z_i+J_2S^z_j+J_3S^z_iS^z_j)+\\
&+&\Lambda \sum_{i<j}\frac{S^z_iS^z_{j}}{(j-i)^3}+\frac{U}{2}\sum_i\left(S^{z}_{i}\right)^{2}-k\sum_i(i-L/2)^2S^z_i \nonumber
\end{eqnarray}
where the first line includes a nearest-neighbor exchange with $J=\alpha^2-\beta^2$ and correlated exchange terms with $ J_1=\sqrt{2}\beta, \;J_2=2\beta$ and $J_3=4\beta^2$, which break particle-hole symmetry, as required for constrained bosons.

We notice that  Eq. (\ref{hamspin}) is a generalization of  the so called $\lambda-D$ model\cite{affleck}, extensively studied over the last two decades both from  analytical and  numerical points of view. 

The spin-$1$ $\lambda-D$ model presents a rich phase diagram: in addition to ferromagnetic and antiferromagnetic (AFM) phases, in a broad region of the parameter space competition between local and non-local interactions favors  the so-called \emph{Haldane phase} (as expected for integer spin chains\cite{haldane}), which displays a gap in the energy spectrum, a unique ground state (at least in  the thermodynamic limit, whereas it is four-fold degenerate for chains of finite size), a finite correlation length, and thus no long-range order even if it is possible to define suitable string correlation functions that measure a hidden topological order. The spin liquid picture introduced by Tasaki\cite{tasaki} provides a intuitive understanding of the Haldane phase: let us assign the presence of an effective spin-$1/2$ particle with spin pointing up (down) if at the $i$-th lattice site $S^{z}_{i} = +1 (-1)$ and no particles if $S^z_{i} = 0$. The Haldane phase is then interpreted as a liquid in which these effective particles carry no positional order along the chain but still retain antiferromagnetic (AFM) order in their effective spins. The positional disorder is associated with the absence of long-range order in the usual spin-1 correlation functions
\begin{eqnarray}\label{SzSz}
\mathcal{C}_{\alpha}(j)=(-1)^{j}\langle S^{\alpha}_{i}S^{\alpha}_{i+j}\rangle\hspace{1cm}\alpha=x,y,z
\end{eqnarray}
whereas the spin-$1/2$ magnetic order that we would get if all the sites with $S^{z}_{i} = 0$ were taken off from the chain is measured by the asymptotic value of the string order parameters (SOP)\cite{dennijs}:
\begin{eqnarray}\label{string}
\mathcal{O}_{\alpha}(j)=\langle S^{\alpha}_{l}e^{i\pi\sum_{l<k<j+l}S^{\alpha}_{k}}S^{\alpha}_{l+j}\rangle,	\quad\alpha=x,y,z.
\end{eqnarray}
As shown thoroughly by Kennedy and Tasaki \cite{kennedy} the $\lambda-D$ model possesses an hidden (non-local) $Z_{2}\times Z_{2}$ symmetry, and the non-vanishing values of the SOP can be understood as the breaking of such a symmetry.

\section{Homogeneous phase diagram.}\label{hom_section}

 \begin{figure}[tb]{
\begin{center}{
\includegraphics[width=8cm]{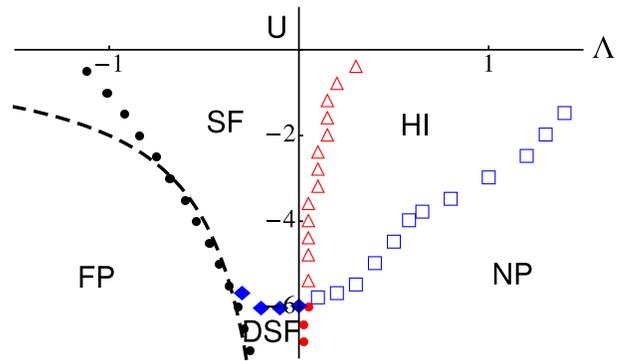}
\caption{(color online): homogeneous phase diagram for dipolar bosons on an optical lattice with three-body hard core constraint at filling $\bar{n}=1$ (see text): triangles, squares, diamonds, black and red points denote numerical results, while the black dashed line describes an approximate strong coupling description for $|U|\gg J$.}
 \label{hom_phase_diagram}}
 \end{center}}
\end{figure}

\begin{figure}[t]{
\begin{center}{
\includegraphics[width=5cm]{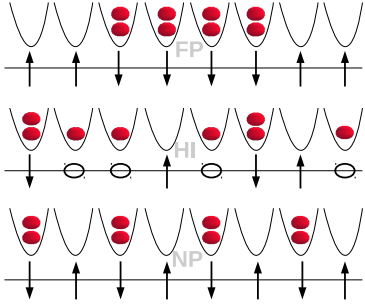}
\caption{(color online): cartoon of magnetic phases related to the model in Eq.(\ref{boseham},\ref{hamspin}): from top to bottom, ferromagnetic, Haldane and N\'eel phase in bosonic and spin language(see text).}
 \label{cartoons}}
 \end{center}}
\end{figure}

In order to exploit a complete parallelism between Eq. (\ref{hamspin}) and the $\lambda-D$ model, we investigate its phase diagram in the homogeneous case, $k=0$, by means of numerical simulations based on the density-matrix renormalization group (DMRG) algorithm \cite{white}, truncating the dipolar interaction up to fifth-nearest-neighbors \cite{comment2}.  
Let us summarize the main results, as schematically presented in Fig. \ref{hom_phase_diagram}: the $\Lambda>0$ region displays {\it i)} an antiferromagnetic N\'{e}el-like phase (NP), where doubly occupied sites alternate with empty ones in a periodic pattern, {\it ii)} an Haldane insulator phase (HI), where doubly occupied and empty sites are separated by strings of single occupied ones\cite{dallatorre,Nonne}(see Fig.  \ref{cartoons}), and {\it iii)} two superfluid phases, in which the superfluid components are single bosons (SF) and dimers (DSF) respectively. In the $\Lambda<0$ regime, both superfluid phases collapse beyond a critical value of $\Lambda$ into a ferromagnetic phase (FP), where the mutual attraction between bosons gives rise to a region of constant density $\bar{n}=2$. The system thus displays all phases and phase transitions of the $\lambda-D$ model with attractive single-ion anisotropy; there are however some quantitative differences. First, both SF and DSF, which correspond to the XY phases in spin language, extend on a broad region around $\Lambda=0$ due to the presence of correlated hopping terms which disadvantage long-range  order. In addition, the HI region is present even at larger $\Lambda$, as expected due to long-range frustration of dipolar interactions with respect to antiferromagnetic ordering\cite{dallatorre}.

\begin{figure}[t]{
\begin{center}{
\includegraphics[width=7.5cm]{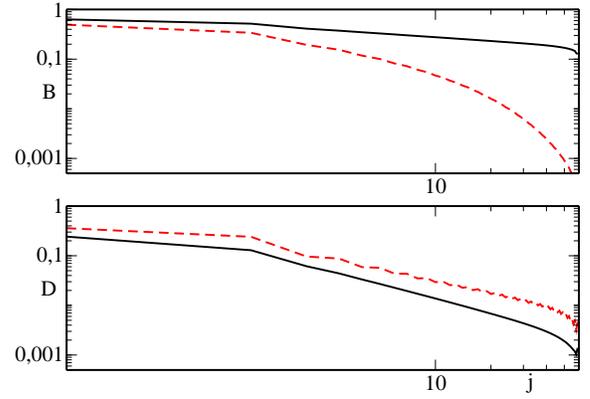}
\caption{(color online): superfluid correlations in double logarithmic scale as a function of the distance from the middle in a $L=120$ chain. Red (dashed) and black (thick) lines represent  DSF and SF phase respectively, with $U=-6.5, \Lambda=0$ and $U=-3, \Lambda=0.05$.
}
 \label{SF_DSF}
 }\end{center}
 }
\end{figure}

\begin{figure}[b]{
\begin{center}{
\includegraphics[width=7.5cm]{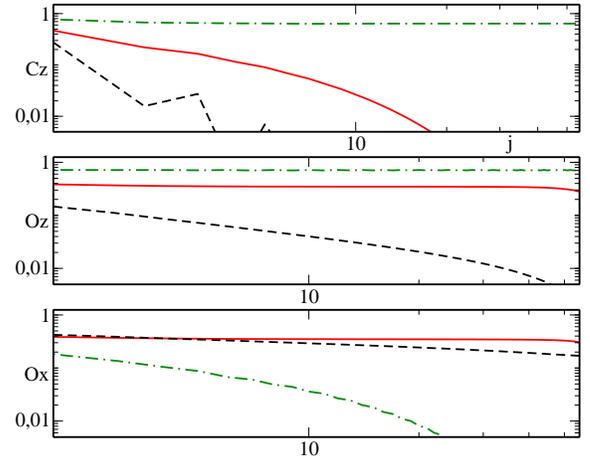}
\caption{(color online): magnetic order parameters in double logarithmic scale as a function of the distance from the middle in a $L=120$ chain.        Black (dashed), red (thick) and  green (dot-dashed) represent $(U=-3,\Lambda=0.05), (-3, 0.7), (-3, 1.3)$ respectively. }
 \label{corr_funct}
 }\end{center}
 }
\end{figure}

\begin{figure}[t]{
\begin{center}{
\includegraphics[width=8.4cm]{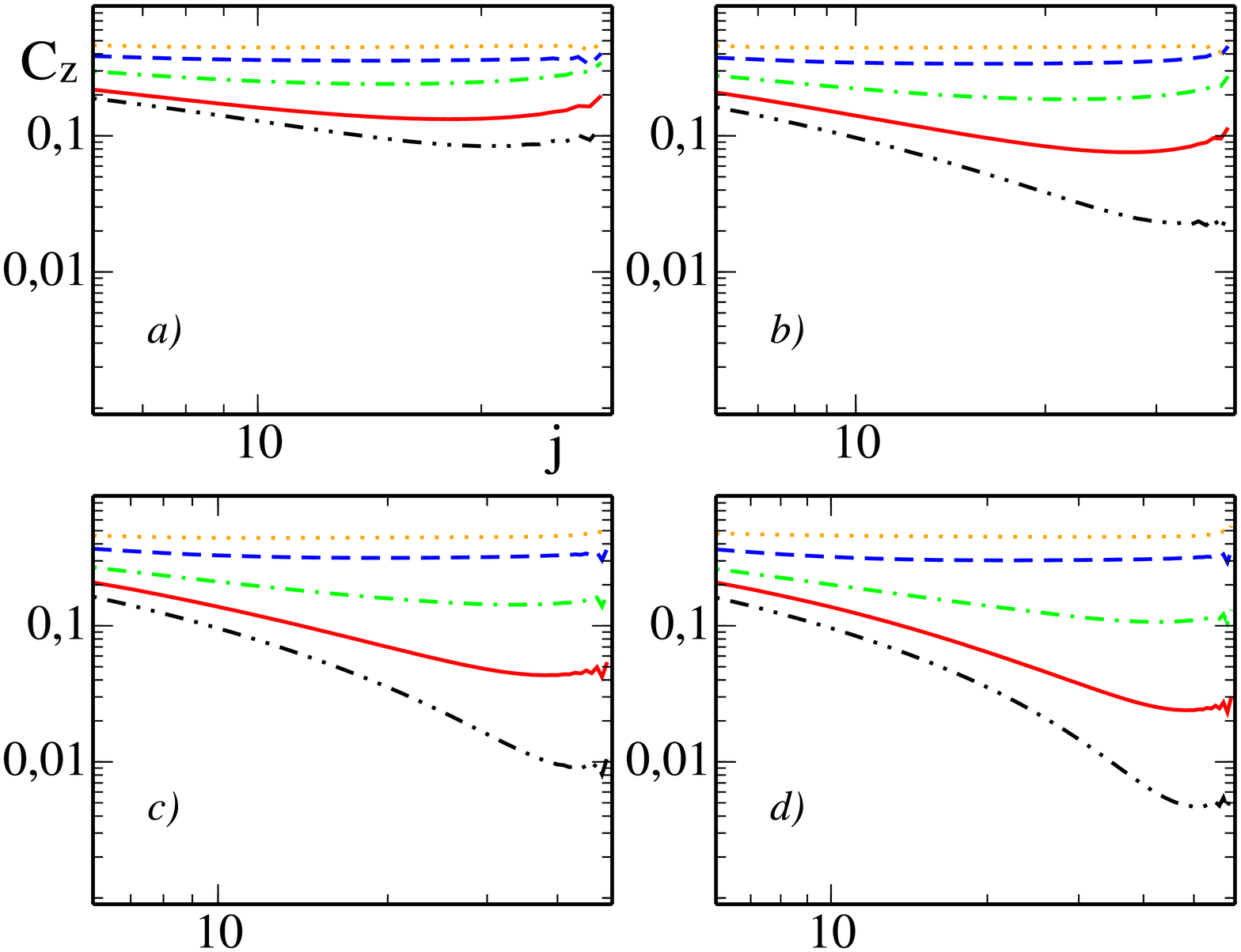}

\caption{(color online).  $\mathcal{C}_z(x)$ correlation function at the HI-NP boundary for  chains of different lengths $L=60, 80, 100, 120$, panels {\it a,b,c} and {\it d} respectively; correlations are taken with respect to the center of the chain. Here, $U=-3$, and, from top to bottom, $\Lambda=1.1$ (orange, dotted), $\Lambda=1.05$ (blue, dashed), $\Lambda=1$ (green, dot-dashed), $\Lambda=0.95$ (red, thick) and $\Lambda=0.9$ (black, dot-dot-dashed).}
 \label{np_hi_corr}
 }\end{center}
 }
\end{figure}

\begin{figure}[t]{
\begin{center}{
\includegraphics[width=8.4cm]{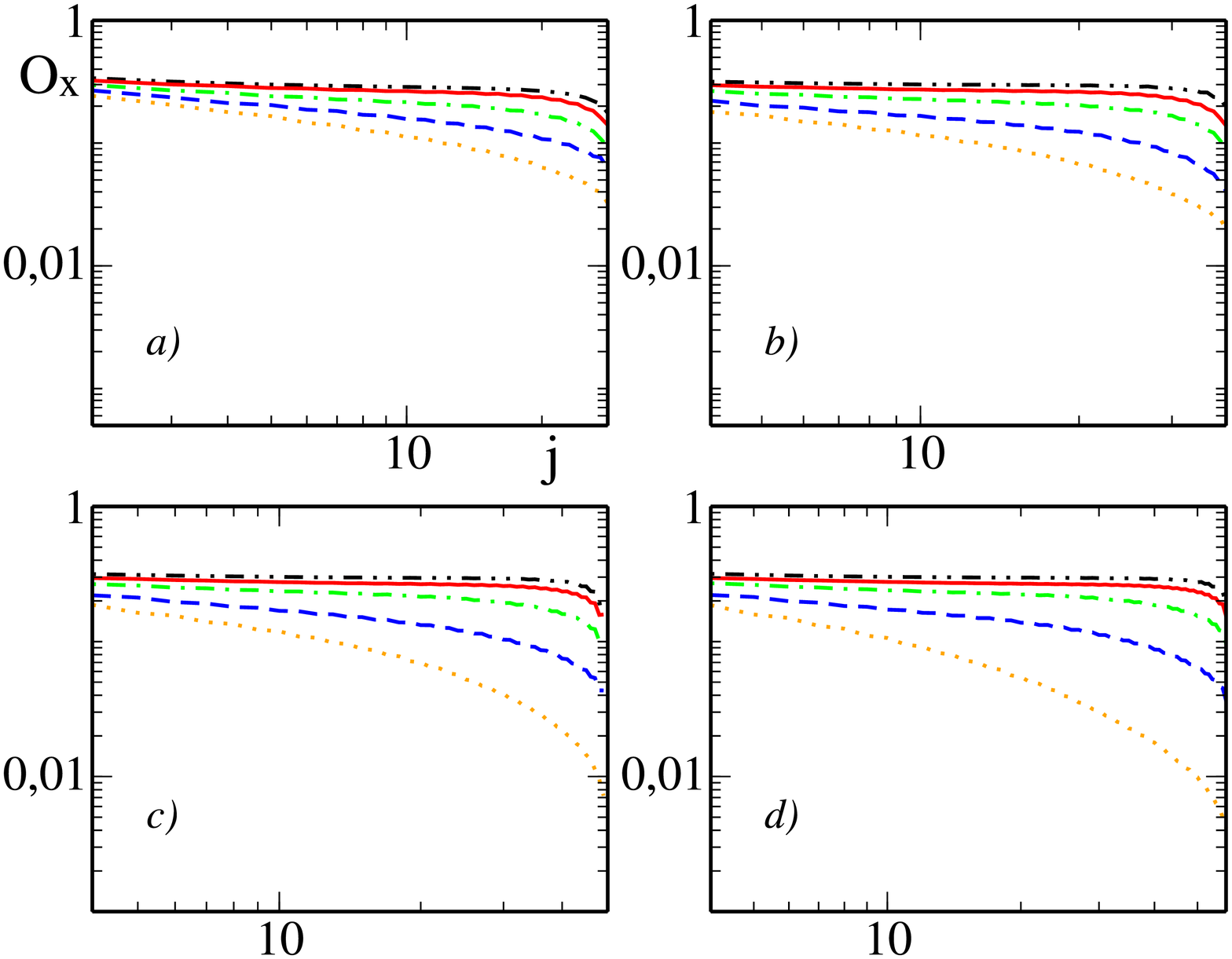}

\caption{(color online). String correlator $\mathcal{O}_x(x)$  at the HI-NP boundary for  chains of different lengths $L=60, 80,100, 120$, panels {\it a,b,c} and {\it d} respectively; correlations are taken with respect to the center of the chain. Here, $U=-3$, and, from top to bottom, $\Lambda=0.9$ (black, dot-dot-dashed), $\Lambda=0.95$ (red, thick), $\Lambda=1$ (green, dot-dashed), $\Lambda=1.05$ (blue, dashed) and $\Lambda=1.1$ (orange, dotted).}
 \label{np_hi_corr2}
 }\end{center}
 }
\end{figure}

Different phases are uniquely characterized by the asymptotic decay of correlation functions\cite{schulz}. In the SF phase, both single particle and dimer superfluid correlations
\begin{equation}
\mathcal{B}(j)=\langle b^{\dagger}_i b_{i+j}\rangle\propto \mathcal{C}_x(j), \quad \mathcal{D}(j)=\langle (b^{\dagger}_i)^2 (b_{i+j})^2  \rangle
\end{equation}
decay algebraically; by contrast, in all other phases $\mathcal{B}$ is exponentially suppressed, whereas $\mathcal{D}$ decays algebraically in the DSF phase, as can be seen in Fig. \ref{SF_DSF}. Magnetic phases are instead characterized by a non-vanishing asymptotic value of certain correlation functions: in the HI, both $\mathcal{O}_x,\mathcal{O}_z$ decay to a constant at long distances  while $\mathcal{C}_{z}$ vanishes exponentially, whereas in the NP $\mathcal{O}_x$ decays exponentially and $\mathcal{O}_z,\mathcal{C}_z$ are constant. All magnetic order parameters decay at long distances in both SF and DSF phases. 
Correlation functions have been computed by analyzing systems of size L=60, 80, 100 and 120 sites, with up to 600 states per block, 4 sweeps and open boundary conditions. Fig. \ref{corr_funct} describes typical decays in the SF (black, dashed), HI (red, thick) and NP (green, dot-dashed) of the magnetic order parameters. 
\paragraph*{Haldane insulator - N\'{e}el phase}. Any one of the pairs $\left\{\mathcal{C}_{z}, \mathcal{O}_{z}\right\}$, $\left\{\mathcal{C}_{z}, \mathcal{O}_{x}\right\}$ or  $\left\{\mathcal{O}_{x}, \mathcal{O}_{z}\right\}$ can be used to give an accurate description of the HI-NP boundary: 
from Figs. \ref{np_hi_corr}, \ref{np_hi_corr2}, it can be inferred that the bulk asymptotic behavior of these correlators is well described already for $L=60$. This feature is not surprising, considering that in the $\lambda-D$ model this transition is believed to belong to the Ising-type universality class\cite{sanctuary}. However, the same cannot be said when considering the other phase transitions present in the model.

\begin{figure}[t]
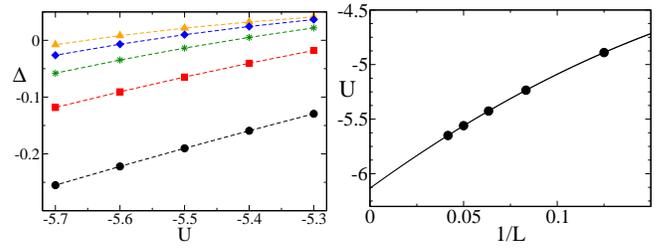
{
\begin{center}{
\includegraphics[width=4.2cm]{Fig7a}
 \includegraphics[width=4.2cm]{Fig7b}

\caption{(color online). {\it Left panel}: dependence of $\Delta=\Delta_1-\Delta_2$ on $U$ for $\Lambda=-0.1$ and different chain lengths: $L=8$ (black circles), $L=12$ (red squares), $L=16$ (green stars), $L=20$ (blue diamond) and $L=24$ (orange triangles). Dashed lines are guides for the eye. {\it Right panel}: critical value of the SF-DSF transition as a function of $1/L$ for $\Lambda=-0.1$; circles represent numerical datas, line is a best fit of the type $a_1+a_2/L+a_3/L^2$.}
 \label{delta_vs_U}
 }\end{center}
 }
\end{figure}

\paragraph*{Superfluid - Dimer superfluid}. The SF-DSF phase transition corresponds to a level crossing in the spectrum between excitations with $S^z=\pm1$ and $S^z=\pm2$, with finite size gaps $\Delta_1$ and $\Delta_2$ respectively \cite{sanctuary}. This condition stems from the fact that, beyond a critical attraction $-U_c(\Lambda)\gg J$, a finite energy is required to break \emph{dimers}, and thus exciting the system in the $S^z=\pm1$ sector would become energetically unfavorable. We determine the phase boundary (marked by blue diamonds in Fig. \ref{hom_phase_diagram}) by calculating finite size gaps 
\begin{equation}
\Delta_{\alpha=1,2}(L)=\frac{\mathcal{E}(N+\alpha; L)+\mathcal{E}(N-\alpha; L)-2\mathcal{E}(N; L)}{2}\nonumber
\end{equation}
 for periodic chains of several lengths and then by imposing that, at the phase transition, the condition $\lim_{L\rightarrow\infty}(\Delta_1(L)-\Delta_2(L))=0$ is satisfied; a typical set of data for $\Lambda=-0.1$ is presented in Fig. \ref{delta_vs_U}.

\begin{figure}[b]{
\begin{center}{
\includegraphics[width=7.2cm]{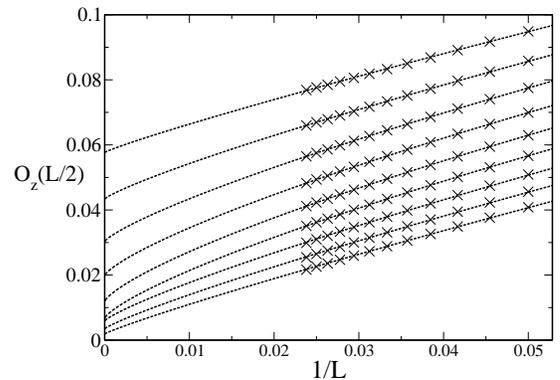}
\caption{Finite-size scaling of $\mathcal{O}_z(L/2)$ for $U=-0.8$ by using Eq. (\ref{C1_fit}) (see text). From top to bottom: $\Lambda=0.5, 0.45, 0.4, 0.35, 0.3, 0.25, 0.2, 0.15$ and $0.1$.}
 \label{Oz_scaling}
 }\end{center}
 }
\end{figure}

\paragraph*{Superfluid - Haldane insulator and Dimer superfluid - N\'{e}el phase}. Finite size calculations are also useful to better shape the SF-HI and DSF-NP transitions, which, in analogy with the $\lambda-D$ model\cite{alcaraz, cristian, ueda}, should belong to the Berezinskii-Kosterlitz-Thouless (BKT) universality class\cite{sachdev}, albeit the non trivial nature of non-local interaction can in principle lead to different critical behaviors. BKT transitions are usually hard to determine due to the exponential opening of the gap; however, string order parameters have been shown to provide a rather accurate estimate of the transition points\cite{alcaraz, ueda}. In the following, we consider systems with   
periodic boundary conditions, in order to avoid boundary effects, with up to $L=42$ sites, and calculate the string order parameter from the first to the $L/2+1$ site, $\mathcal{O}_z(x-x'=L/2)$, which represents a suitable order parameter for both SF-HI and DSF-NP transitions\cite{sanctuary}. Then, we estimate its asymptotic value $C_1$ by fitting the data with the following scaling form:
\begin{equation}\label{C1_fit}
\mathcal{O}_z(L/2)=C_1+\frac{C_2}{L^{C_3}}.
\end{equation}
\begin{figure}[t]{
\begin{center}{
\includegraphics[width=6.5cm]{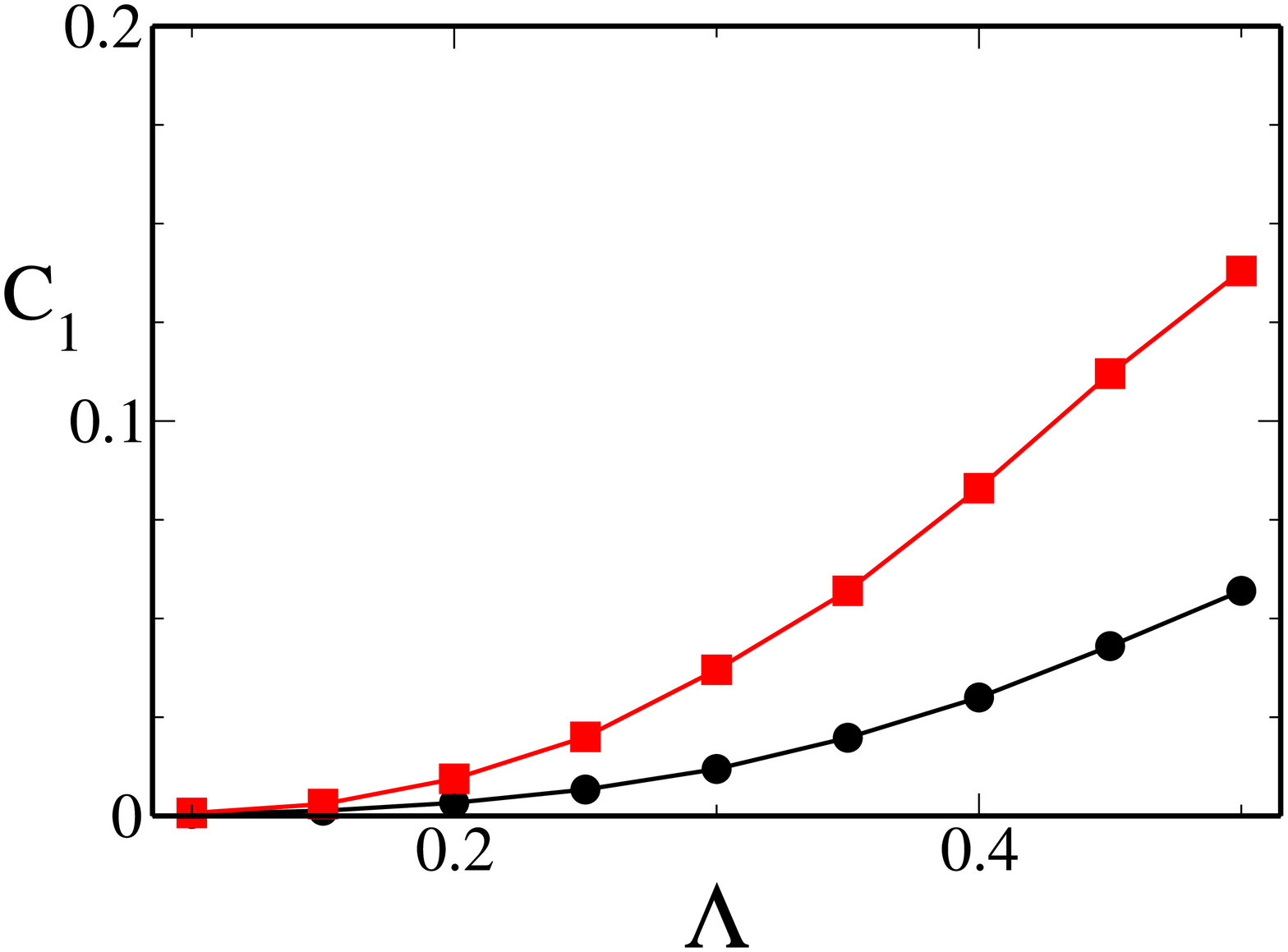}
% \vspace{1.5cm}
 %\bigskip
 \includegraphics[width=6.5cm]{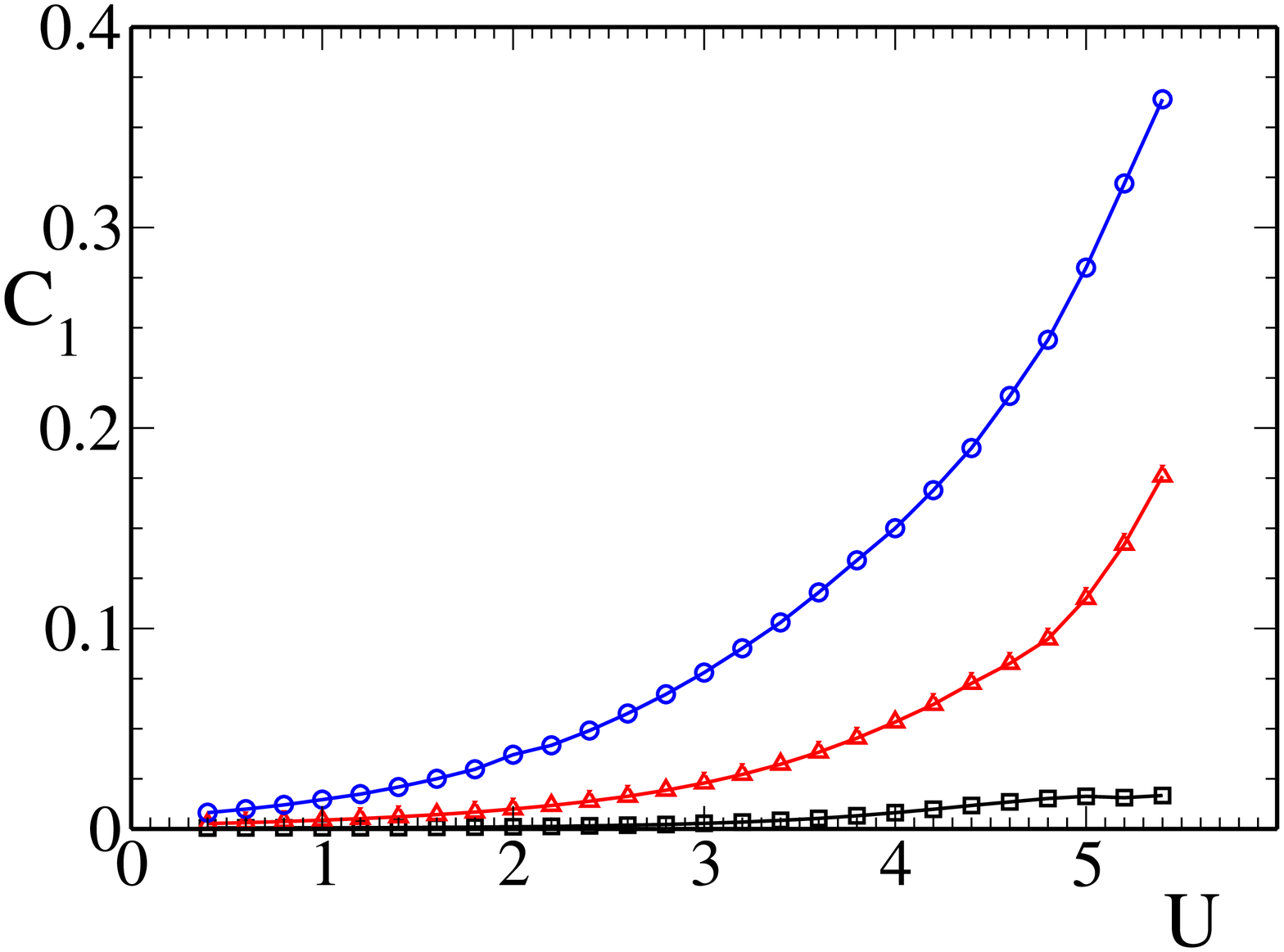}
\caption{(color online). {\it Upper panel}: string order parameter $C_1$ for $U=-0.8, -2$ (black circles and red squares respectively) as a function of $\Lambda$. {\it Lower panel}: $C_1$ for $\Lambda=0.1,0.2,0.3$ (black squares, red triangles and blue circles respectively) as a function of $U$. In both panels, lines are guide for the eye, and the size of each point denotes the maximum error on the extrapolated value $C_1$.}
 \label{C1}
 }\end{center}
 }
\end{figure} 
This method has been successfully employed to study the same transition in the $\lambda-D$ model within an exact diagonalization approach\cite{alcaraz}.
The usual error of this procedure is related to the DMRG truncation error, always smaller than $3*10^{-5}$, and to the algebraic fit \cite{numrecipe}: by employing numerical datas with $20\leq L\leq42$, we then estimate that the asymptotic value $C_1$ is non-vanishing within numerical error as long as $C_1>0.005$. 
 A typical example of the estimate of $C_1$ is described in Fig. \ref{Oz_scaling}, where different datas are presented for $U=-0.8,0.1\leq\Lambda\leq0.5$. In Fig. \ref{C1}, we plot $C_1$ as a function of both $U$ (fixing $\Lambda=0.1, 0.2, 0.3$) and $\Lambda$ (fixing $U=-0.8, -2$); the asymptotic value of the string order parameter increases with both $\Lambda$ and $|U|$. The corresponding transition points are 
marked by red triangles in Fig. \ref{hom_phase_diagram}; due to the small system sizes analyzed here, the numerical data presented above can be considered as an approximate estimate on the BKT transition points, an accurate one needing more specific techniques\cite{ueda, sanctuary}.

\paragraph*{Ferromagnetic phase - Superfluid}. A clear evidence of the FP-SF transition emerges instead when the ground state energy density $\mathcal{E}_{GS}(L)$ approaches the exact value for ferromagnetic states $\mathcal{E}_{FP}$ in the thermodynamic limit\cite{sanctuary}. The transition line, marked by black dots in Fig.(\ref{hom_phase_diagram}), is obtained by requiring that $\lim_{L\rightarrow\infty}\mathcal{E}_{GS}(L)=\mathcal{E}_{FP}$.

\subsection{Strong coupling regime}
In the large $|U|$ regime, the quantitative difference between dipolar and nearest-neighbor(NN) interaction can be investigated with a perturbative argument. If $|U|\gg |\Lambda|, J$, the effective Hilbert space is reduced to $S^z=\pm1$, that is, bosons are tightly bound in dimers, so that we can map the spin-1 problem into a spin-$1/2$ theory employing the following identities\cite{cristian}:
\begin{equation}
S_{i}^{z}=2s_{j}^{z},\qquad S_{j}^{+}S_{j}^{+}=2s_{j}^{+},\qquad S_{j}^{-}S_{j}^{-}=2s_{j}^{-}
\end{equation}
\begin{equation}
S_{j}^{+}S_{j}^{-}=2(1/2+s_{j}^{z}),\qquad S_{j}^{-}S_{j}^{+}=2(1/2-s_{j}^{z})
\end{equation}
where $\vec{s}$ is a spin-$1/2$ operator. After a proper rescaling, the strong coupling Hamiltonian including only NN interaction is mapped into a spin-$1/2$ XXZ chain: 
\begin{equation}\label{XXZ}
H_{sc}=\sum_{<i,j>}(s^+_is^-_j+(1-\Delta)s^z_is^z_j), \quad \Delta=-2/\Lambda|U|.
\end{equation}
From the exact solution of Eq.(\ref{XXZ})\cite{korepin}, we argue that the system is in a DSF phase as long as $ -\frac{2}{|U|} \leq\Lambda\leq 0$; the DSF-FP and DSF-NP transitions are located at $\Lambda_{NN}^{(c)}=-\frac{2}{|U|},0$ respectively. We can now compare this criterion, derived considering only NN interactions, with the numerical one, obtained from DMRG as previously described. 
The DSF-FP transition (black dashed line in Fig.\ref{hom_phase_diagram} ) is then predicted at $-\Lambda\simeq 2/|U|$, whereas numerical values (including dipolar interaction) for, e.g., $U=-10$, indicate $-\Lambda\simeq 1.75/|U|$; in this regime, dipolar interactions show a small quantitative difference with respect to standard NN couplings.

\section{Effect of a trapping potential.}\label{trap}
The observation of the different magnetic orders discussed above in a standard cold atom experiment is strictly related to the possibility of stabilizing these phases even in an inhomogeneous background. In fact, atoms and molecules are loaded into a trapping potential, which introduces a position-dependent term in the Hamiltonian with a minimum at the trap center, namely, the first term in the second line of Eq.(\ref{boseham}). In this setup, another energy scale comes into play; particles will try to minimize their potential energy by concentrating in the middle of the trap, thus displaying a strong spatial dependence of the local density $\langle n_i\rangle$. This feature is in sharp contrast to the ideal configuration needed to realize the magnetic phases described above, all of them requiring a constant density in the thermodynamic limit.
Our goal here is to investigate what are the proper trap configurations needed to stabilize a magnetic phase in this inhomogeneous setup: in particular, we will focus our attention to the region in the middle of the confining potential, where it is usually easier to create large regions of space at constant density\cite{bloch_rv, bakr}. First of all, let us briefly discuss what happens the in the FP:  in analogy to a standard Mott-like phase\cite{bloch_rv}, it can always be realized by considering a sufficiently strong trap such that the density is maximized in the middle,  $\langle n_i\rangle=2$. This simple argument cannot be extended to neither HI or NP: in fact, a very strong trap will simply destroy these types of order. In this section, we will thus focus on the stability of these two orders in presence of a trapping potential. 

We identify a certain magnetic order in a region of space by requiring that {\it i)}  the region is at constant density, $\langle n_i\rangle=1$, and {\it ii)} the proper order parameters with respect to the middle of the trap, defined as:
\begin{eqnarray}\label{stringtrap}
\mathcal{O}_{\alpha=x,z}(L/2,j)=\langle S^{\alpha}_{L/2}e^{i\pi\sum_{L/2<k<j+L/2}S^{\alpha}_{k}}S^{\alpha}_{L/2+j}\rangle\quad
\end{eqnarray}
\begin{eqnarray}\label{SzSztrap}
\mathcal{C}_{z}(L/2,j)=(-1)^{j}\langle S^{z}_{L/2}S^{z}_{L/2+j}\rangle
\end{eqnarray}
where $j$ is the distance from $L/2$,  behave as expected in the HI or in the NP up to a certain range. We performed DMRG simulations  on a $L=80$ sites chain, fixing as energy unit for the trapping potential $k^*=1/(L/2)^2=1/1600$, and kept as much as 600 states per block with 10 finite-size sweeps\cite{white}.  Since a trapping potential favors a configuration where the particles are in the middle of the chain, we focused on a $N=40$ particles system:  this assures that the density close to the chain boundary rapidly goes to zero, thus avoiding possible finite-size effects and, at the same time, allows for a constant density of order 1 in the middle of the system.

\begin{figure}[t]{
\begin{center}{
\includegraphics[width=6.5cm,angle=-90]{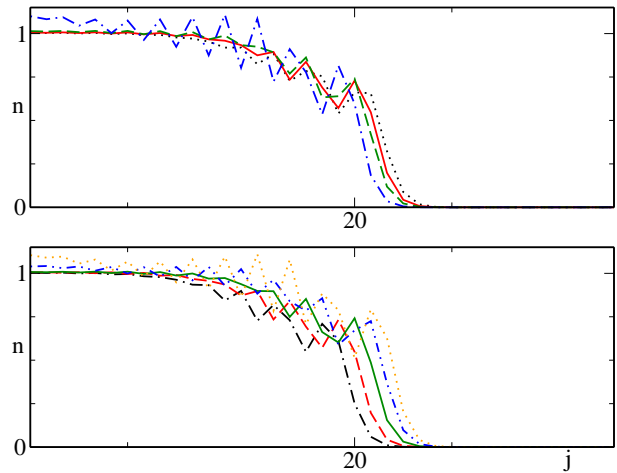}
\caption{(color online): density distribution as a function of the distance from the trap center for $U=-2.5, \Lambda=0.9$. {\it Top panel}: fixed population $N=40$ and different trap strength $k=7.5k^*$ (black, dotted), $8.5k^*$ (red, thick), $9.5k^*$ (green, dashed) and $10.5k^*$ (blue, dot-dashed). {\it Bottom panel}: fixed trap strength $k=9k^*$ and different populations: N=36 (black, dot-dashed), 38 (red, dashed), 40 (green, thick), 42 (blue, dot-dot-dashed) and 44 (orange, dotted).}
 \label{haldane_trap}
 }\end{center}
 }
\end{figure}

\subsection{Haldane order}
 We start our treatment by considering the possibility to stabilize hidden order in an inhomogeneous  system. As a sample configuration, we fixed $U=-2.5, \Lambda=0.9$ such that the corresponding homogeneous phase at integer filling is an HI. In a very shallow trap, the non-local interparticle repulsion would drive the system in a very dilute limit with $\langle n_i \rangle<1$ all over the trap, whereas in the opposite strong trap limit, an high density region with $\langle n_i\rangle>1$ would be stabilized in the middle. We shall then focus on an intermediate regime in order to satisfy the density requirement $\langle n_{i}\rangle= 1$ needed in the HI.

\begin{figure}[t]{
\begin{center}{
\includegraphics[width=7cm]{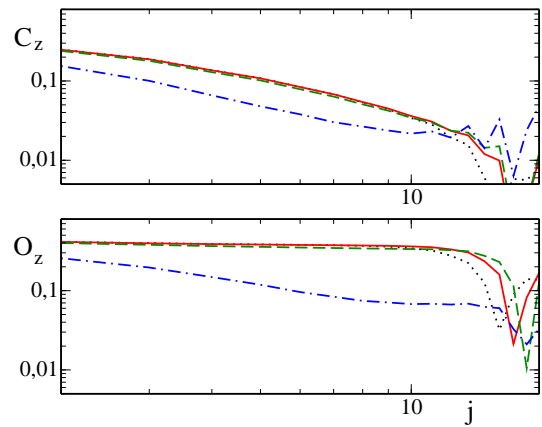}
\caption{(color online): magnetic order parameters $\mathcal{C}_z(L/2,j)$ (top panel) and  $\mathcal{O}_z(L/2,j)$ (bottom panel) as a function of the distance from the trap center for $U=-2.5, \Lambda=0.9, N=40$ and different values of $k$: $k=7.5k^*$ (black, dotted), $8.5k^*$ (red, thick), $9.5k^*$ (green, dashed) and $10.5k^*$ (blue, dot-dashed).}
 \label{haldane_trap_k}
 }\end{center}
 }
\end{figure}

\begin{figure}[t]{
\begin{center}{
\includegraphics[width=7cm]{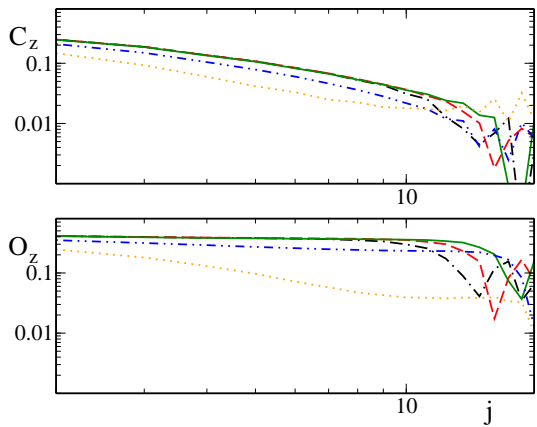}
\caption{(color online): magnetic order parameters $\mathcal{C}_z(L/2,j)$ (top panel) and  $\mathcal{O}_z(L/2,j)$ (bottom panel) as a function of the distance from the trap center for $U=-2.5, \Lambda=0.9, k=9k^*$ and different number of particles N=36 (black, dot-dashed), 38 (red, dashed), 40 (green, thick), 42 (blue, dot-dot-dashed) and 44 (orange, dotted). }
 \label{haldane_trap_L}
 }\end{center}
 }
\end{figure}

In Fig. \ref{haldane_trap}, upper panel, we plot the density distribution as a function of the distance from the trap center for different  values of  $k$, $7.5\leq k/k^*\leq10.5$:  the requirement $\langle n_{i}\rangle= 1$ is satisfied for values of the trap strength inside the interval $8.5\leq k/k^*\leq9.5$. In order to verify whether hidden order is present or not, we plot the relevant magnetic order parameters as defined in Eqs. \ref{stringtrap}, \ref{SzSztrap} in Fig. \ref{haldane_trap_k}; in the interval $8.5\leq k/k^*\leq9.5$, the string order parameter $\mathcal{O}_{z}$ is constant up to a certain distance from the trap, and at the same time $\mathcal{C}_z$ decays, then proving that particles close to the trap center display HI; outside of the constant density region, the order is lost, as can also be seen by looking at $\mathcal{O}_z$.

However, while an accurate fine tuning of the trap strength does not present major difficulties in a typical experimental setup, a proper control over populations in a tube is challenging, and it is thus instructive to investigate small population unbalance with respect to the previous $N=40$ case. In Fig. \ref{haldane_trap}, lower panel, we plot the density distribution at a fixed trap strength $k=9k^*$ for different total number of particles $N=36,38,40,42$ and 44, while the corresponding order parameters are plotted in Fig. \ref{haldane_trap_L}. We notice that the Haldane phase is unstable when $N\geq42$ since too many particles concentrate in the middle of the trap, whereas it is stable for $N\leq 40$; we can then conclude that a large population difference of order $\delta N\sim 0.1$ prevents the HI phase to stabilize in the center of the trap.

\begin{figure}[t]{
\begin{center}{
\includegraphics[width=8cm]{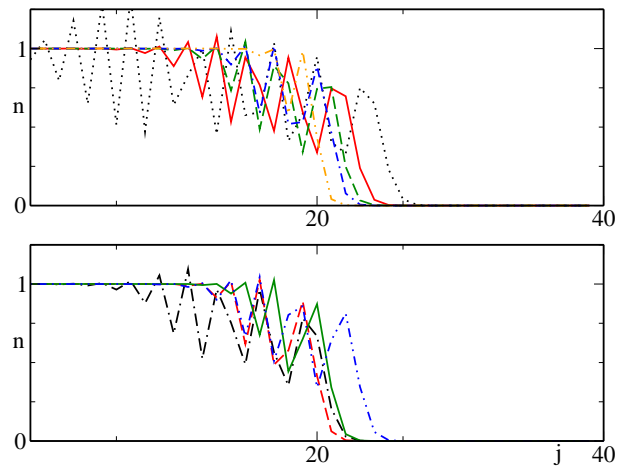}
\caption{(color online): density distribution as a function of the distance from the trap center for $U=-4, \Lambda=1.5$. {\it Top panel}: fixed population $N=40$ and different trap strength $k=5k^*$ (black, dotted), $7k^*$ (red, thick), $9k^*$ (green, dashed), $11k^*$ (blue, dot-dashed) and $20k^*$ (orange, dot-dot-dashed). {\it Bottom panel}: fixed trap strength $k=12k^*$ and different populations: N=36 (black, dot-dashed), 38(red, dashed), 40 (green, thick) and  42 (blue, dot-dot-dashed).}
 \label{neel_trap}
 }\end{center}
 }
\end{figure}

\subsection{Antiferromagnetic ordering}

\begin{figure}[t]{
\begin{center}{
\includegraphics[width=7 cm]{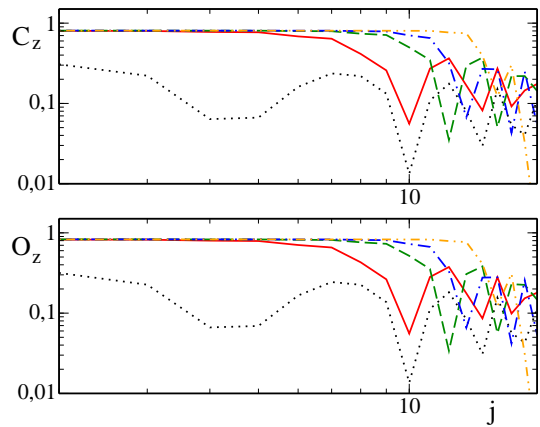}
\caption{(color online): magnetic order parameters $\mathcal{C}_z(L/2,j)$ (top panel) and  $\mathcal{O}_z(L/2,j)$ (bottom panel) as a function of the distance from the trap center for $U=-4, \Lambda=1.5, N=40$ and different values of $k$: $k=5k^*$ (black, dotted), $7k^*$ (red, thick), $9k^*$ (green, dashed), $11k^*$ (blue, dot-dashed) and $20k^*$ (orange, dot-dot-dashed).}
 \label{neel_trap_k}
 }\end{center}
 }
\end{figure}

We turn now our attention to the NP by fixing $U=-4, \Lambda=1.5$. As already discussed for the HI, a very shallow trap is not sufficient to stabilize antiferromagnetic order in the trap due to density requirements, whereas a too strong trap would prevent it by concentrating too many particles in the trapping potential minimum. The density distribution as a function of the trap strength in the interval $5\leq k/k^*\leq 20$ is presented in Fig. \ref{neel_trap}, upper panel; a large region with $\langle n_i\rangle=1$ is stable in the middle as long as $k>5k^*$, and, remarkably, the size of this region increases with increasing $k$, including up to 30 particles when $k=20k^*$. The corresponding magnetic order parameters are plotted in Fig. \ref{neel_trap_k}; both $\mathcal{C}_z$ and $\mathcal{O}_z$ are constant in the middle of the trap as long as $k> 5k^*$, and their plateau extends all over the constant density region.

\begin{figure}[t]{
\begin{center}{
\includegraphics[width=7 cm]{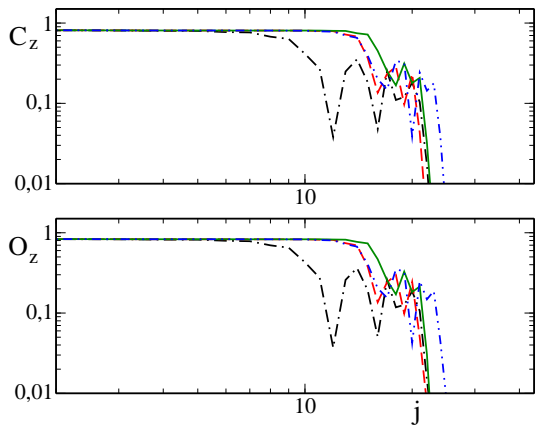}
\caption{(color online): magnetic order parameters $\mathcal{C}_z(L/2,j)$ (top panel) and  $\mathcal{O}_z(L/2,j)$ (bottom panel) as a function of the distance from the trap center for $U=-4, \Lambda=1.5, k=12k^*$ and different number of particles: N=36 (black, dot-dashed), 38(red, dashed), 40 (green, thick) and  42 (blue, dot-dot-dashed). }
 \label{neel_trap_L}
 }\end{center}
 }
\end{figure}

Small changes in the total number of particles do not alter this picture significantly. In Fig. \ref{neel_trap}, lower panel, we plot the density distribution at a fixed trap strength $k=12k^*$ and different total number of particles $N=36,38,40,42$; a constant region in the middle of the trap is always present and, in addition, it displays antiferromagnetic correlations, as can be inferred from the magnetic order parameters presented in Fig. \ref{neel_trap_L}.

We can then conclude that, while both the HI and NP are incompressible, the former requires a finer tuning of the trapping potential  and a more accurate control of the population of the system in order to be stabilized in the centre of the trap.

\section{Conclusions}\label{conclusions}

We have investigated the effect of a three-body hard-core constraint in a one dimensional system of dipolar bosons such as magnetic atoms or polar molecules confined in a one dimensional tube by optical lattices. After mapping the original problem in spin language, a strong parallelism between the system and the $\lambda-D$ model is established and confirmed by DMRG calculations; we have shown that ferromagnetic, antiferromagnetic and hidden orders can be stabilized in this framework, and that  dipolar interactions give rise to small quantitative changes in the phase diagram with respect to more usual nearest-neighbor interactions.

Finally, we have considered the effect of a confining potential, as usually present in cold atomic gas experimental setups. Remarkably, both HI and FP can be stabilized in a large region in the middle of the trap even if the total population is not exactly controlled; the realization of the HI needs however an appropriate trap tuning. This feature opens up the possibility to observe strongly correlated phases in cold gases of magnetic atoms such as \emph{Dy}, \emph{Cr} or \emph{Er}\cite{pfau,pmreviews}, which are usually characterized by relatively small dipolar interactions. All of these phases can be probed via noise correlations\cite{altman}, or, in the HI case, via Bragg spectroscopy\cite{dallatorre} or in-situ imaging\cite{sherson, bakr}.   Finally, this setup can be adapted to investigate spin-1 Heisenberg-like models in 2-D systems, where various interesting phases such as field induced supersolidity have been recently suggested\cite{sengupta}  or else extended in order to consider the effect of disorder in such systems.

\acknowledgements

We acknowledge fruitful discussions with A. J. Daley, E. G. Dalla Torre, C. Degli Esposti Boschi, S. Diehl, E. Ercolessi, P. Lecheminant, A. Micheli, G. Pupillo and P. Zoller. This work is partially supported by Italian MIUR, through the PRIN under Grant No. 2007JHLPEZ.

\appendix

\section{Constrained bosons - spin-1 mapping}\label{app_map}

The constraint $\left(b^{\dag}\right)^{3}|0\rangle=0$ allows us to make the following correspondence between the reduced bosonic Hilbert space and that of a spin-1 
\begin{eqnarray}
|0\rangle\rightarrow |\uparrow\rangle\hspace{1cm}|1\rangle\rightarrow|\tilde{0}\rangle\hspace{1cm}|2\rangle\rightarrow|\downarrow\rangle
\end{eqnarray}
where $|\uparrow\rangle$, $|\tilde{0}\rangle$ and $|\downarrow\rangle$ are eigenstates of $S^{z}$ with eigenvalues $+1$, $0$ and $-1$ respectively. The corresponding operator mapping is:
\begin{eqnarray}
b^{\dag}b=1-S^{z}
\end{eqnarray}
\begin{equation}
b=\alpha S^{+}+\beta\left(S^{z}S^{+}+S^{+}S^{z}\right)
\end{equation}
where the coefficients $\alpha, \beta$ have to be determined by imposing the correct action on the Hilbert space and commutation relations. Verifying the former, we have:
\begin{eqnarray}
\begin{array}{ll}
b|0\rangle=0& \quad\left[\alpha S^{+}+\beta\left(S^{z}S^{+}+S^{+}S^{z}\right)\right]|\uparrow\rangle=0\\
b|1\rangle=|0\rangle&\quad\left[\alpha S^{+}+\beta\left(S^{z}S^{+}+S^{+}S^{z}\right)\right]|\tilde{0}\rangle\\
&\quad=\left(\alpha\sqrt{2}+\beta\sqrt{2}\right)|\uparrow\rangle
\end{array}
\end{eqnarray}
\begin{eqnarray}
\begin{array}{ll}
b|2\rangle=\sqrt{2}|1\rangle&\quad\left[\alpha S^{+}+\beta\left(S^{z}S^{+}+S^{+}S^{z}\right)\right]|\downarrow\rangle\\
&\quad=\left(\alpha\sqrt{2}-\beta\sqrt{2}\right)|\tilde{0}\rangle.
\end{array}
\end{eqnarray}
It follows then
\begin{eqnarray}
\alpha+\beta=\frac{1}{\sqrt{2}}\hspace{2cm}
\alpha-\beta=1.
\end{eqnarray}
Furthermore, if we  write down the number operator in terms of spin-1 operators
\begin{eqnarray}
b^{\dag}b\rightarrow
\mathcal{S}&=&\alpha^{2}S^{-}S^{+}+\beta^{2}\left(S^{-}S^{z}S^{+}S^{z}+S^{-}S^{z}S^{z}S^{+}+\right.\nonumber\\
&+&\left.S^{z}S^{-}S^{+}S^{z}+S^{z}S^{-}S^{z}S^{+}\right)\\
&+& \alpha\beta\left(S^{-}S^{+}S^{z}+2S^{-}S^{z}S^{+}+S^{z}S^{-}S^{+}\right)\nonumber
\end{eqnarray}
and we apply it to number eigenstates, we get:
\begin{eqnarray}
\begin{array}{ll}
b^{\dag}b|0\rangle=0&\mathcal{S}|\uparrow\rangle=0\\
b^{\dag}b|1\rangle=|1\rangle&\mathcal{S}|\tilde{0}\rangle=2\left(\alpha+\beta\right)^{2}|\tilde{0}\rangle=|\tilde{0}\rangle\\
b^{\dag}b|2\rangle=2|2\rangle&\mathcal{S}|\downarrow\rangle=2\left(\alpha-\beta\right)^{2}|\downarrow\rangle=2|\downarrow\rangle.
\end{array}
\end{eqnarray}
We will now show that the operators defined by our mapping satisfy the correct commutation relations. Since we are considering constrained bosons the usual bosonic commutation relation becomes\cite{commentA}
\begin{eqnarray}
\left[b, b^{\dag}\right]=|0\rangle\langle0|+|1\rangle\langle1|-2|2\rangle\langle2|.
\end{eqnarray} 
Recalling that 
\begin{eqnarray}
\left[S^{+},S^{-}\right]&=&2S^{z}\\
\left\{S^{+},S^{-}\right\}&=&2\left(S\left(S+1\right)-\left(S^{z}\right)^{2}\right)\nonumber
\end{eqnarray}
we have
\begin{eqnarray}
\left[b,b^{\dag}\right]=-8\alpha\beta+2\left(\alpha^{2}+\beta^{2}\right)S^{z}+12\alpha\beta\left(S^{z}\right)^{2}\quad
\end{eqnarray}
and then
\begin{eqnarray}
\langle 0|\left[b,b^{\dag}\right]|0\rangle= 2\left(\alpha+\beta\right)^{2}= 1\nonumber
\end{eqnarray}
\begin{eqnarray}
\langle 1|\left[b,b^{\dag}\right]|1\rangle= -8\alpha\beta= 1
\end{eqnarray}
\begin{eqnarray}
\langle 2|\left[b,b^{\dag}\right]|2\rangle= -2\left(\alpha-\beta\right)^{2}= -2.\nonumber
\end{eqnarray}
In order these relations to be satisfied we must have
\begin{eqnarray}
\alpha=\frac{2+\sqrt{2}}{4}\hspace{1.3cm}\beta=-\frac{2-\sqrt{2}}{4}.
\end{eqnarray}


\begin{thebibliography}{39}
\bibitem{comment2}We verified that a simpler nearest-neighbor approximation leads to appreciable quantitative differences in the phase boundaries.
\bibitem{commentA}The factor $-2$ comes from $\langle 2|bb^{\dag}|2\rangle=0$ and not $\langle 2|bb^{\dag}|2\rangle=3$ as in the usual bosonic Hilbert space: thus $\langle 2|\left[b,b^{\dag}\right]|2\rangle=-2$.
\end{thebibliography}
\end{document}